\documentclass[usenatbib]{iau} 
\usepackage{hyperref}
\usepackage{natbib}
\usepackage{aas_macros}
\usepackage{graphicx}

\title[AGN in dwarf galaxies] 
{Feeding and feedback from little monsters: AGN in dwarf galaxies}

\author[Mar Mezcua]   
{Mar Mezcua$^{1,2}$}

\affiliation{$^1$Institute of Space Sciences (ICE, CSIC), Campus UAB, Carrer de Magrans, 08193 Barcelona, Spain \\ email: {\tt marmezcua.astro@gmail.com} \\[\affilskip]
$^2$Institut d'Estudis Espacials de Catalunya (IEEC), Carrer Gran Capit\`{a}, 08034 Barcelona, Spain}

\pubyear{2020}
\volume{359}  
\jname{Galaxy Evolution and Feedback Across Different Environments}
\editors{A.C. Editor, B.D. Editor \& C.E. Editor, eds.}
\begin{document}

\maketitle

\begin{abstract}
Detecting the seed black holes from which quasars formed is extremely challenging; however, those seeds that did not grow into supermassive should be found as intermediate-mass black holes (IMBHs) of 100-10$^5$ M$_{\odot}$ in local dwarf galaxies. The use of deep multiwavelength surveys has revealed that a population of actively accreting IMBHs (low-mass AGN) exists in dwarf galaxies at least out to $z\sim$3. The black hole occupation fraction of these galaxies suggests that the early Universe seed black holes formed from direct collapse of gas, which is reinforced by the possible flattening of the black hole-galaxy scaling relations at the low-mass end. This scenario is however challenged by the finding that AGN feedback can have a strong impact on dwarf galaxies, which implies that low-mass AGN in dwarf galaxies might not be the untouched relics of the early seed black holes. This has important implications for seed black hole formation models. 

\keywords{galaxies: dwarf, active, accretion, nuclei, evolution}
\end{abstract}

\firstsection 
\section{Introduction}
The discovery more than 20 years ago of two low-mass (black hole mass $M_\mathrm{BH} \lesssim10^6$ M$_{\odot}$) active galactic nuclei (AGN; NGC 4395 and POX 52; \citealt{1989ApJ...342L..11F}; \citealt{1987AJ.....93...29K}) triggered a quest that has yielded today more than 500 sources. Most of these low-mass AGN are hosted either by disky (\citealt{2008ApJ...688..159G}; \citealt{2011ApJ...742...68J}; \citealt{2018ApJ...863....1C}) or dwarf galaxies (with stellar mass $M_\mathrm{*} \leq 3 \times 10^{9}$ M$_{\odot}$; \citealt{2013ApJ...775..116R}; Moran et al. 2014) and have been identified based either on narrow emission-line diagnostic diagrams accompanied by the detection of broad emission lines (from which a black hole mass measurement has been obtained; e.g. \citealt{2004ApJ...610..722G,2007ApJ...670...92G}; \citealt{2013ApJ...775..116R}; \citealt{2018ApJ...863....1C}), on high-ionization optical/infrared emission lines (e.g. \citealt{2008ApJ...677..926S}; \citealt{2017A&A...602A..28M}), on the detection of X-ray or radio emission indicative of AGN accretion (e.g. \citealt{2013ApJ...773..150S}; \citealt{2015ApJ...805...12L}; \citealt{2016ApJ...817...20M,2018MNRAS.478.2576M}; \citealt{2019MNRAS.488..685M}; \citealt{2020ApJ...888...36R}), or a combination of all (e.g. see review by \citealt{2017IJMPD..2630021M}; \citealt{2019arXiv191109678G}). A few more tens have been recently identified based on optical variability (e.g. \citealt{2018ApJ...868..152B}; \citealt{2020ApJ...889..113M}). 

The finding of such a number of (low-mass) AGN in dwarf galaxies poses a challenge to galaxy/black hole formation models. How have dwarf galaxies, with their shallow potential well, been able to form a $\sim10^4-10^6$ M$_{\odot}$ black hole at their center? Actually recent studies show that AGN in dwarf galaxies are wandering in their host (e.g. \citealt{2020ApJ...888...36R}), so how have dwarf galaxies been able to assemble an off-nuclear AGN? The answer seems to come from high redshifts ($z\sim$10-20). Low-mass black holes in local dwarf galaxies are thought to be the ungrown relics of the seed black holes formed in the early Universe, with $M_\mathrm{BH}$ ranging from 100 to $\lesssim10^6$ M$_{\odot}$ (e.g. \citealt{2010A&ARv..18..279V,2012Sci...337..544V}; \citealt{2012NatCo...3E1304G}). Such seed black holes have been invoked to explain the finding of quasars hosting supermassive black holes of $10^9-10^{10}$ M$_{\odot}$ by $z\sim$7 (\citealt{2018Natur.553..473B}; \citealt{2019ApJ...872L...2M}) and the presence of overmassive black holes in local brightest cluster galaxies (\citealt{2011Natur.480..215M}; \citealt{2018MNRAS.474.1342M}) and they could have formed from the first Population III stars (e.g. \citealt{2004ARA&A..42...79B}) or from the collapse of metal-free halos and subsequent formation and death of a supermassive star (direct collapse black holes; e.g. \citealt{1994ApJ...432...52L}; \citealt{2013ApJ...778..178H}) among other possible scenarios (see reviews by \citealt{2017IJMPD..2630021M}; \citealt{2019PASA...36...27W}). 

Stellar and supernova (SN) feedback is assumed to be responsible for hampering the growth of the high-$z$ seed black holes via winds that deplete gas from the center (e.g. \citealt{2008MNRAS.383.1079V}; \citealt{2010MNRAS.408.1139V}; \citealt{2017MNRAS.468.3935H}), so that neither the seed black hole nor its host dwarf galaxy grow much through cosmic time and we can observe them today as relics of the first galaxies and first black holes. Recent studies are however starting to challenge this scenario. Both simulations (\citealt{2016MNRAS.463.2986S}; \citealt{2018MNRAS.473.5698D}; \citealt{2019MNRAS.487.5549B}; \citealt{2019MNRAS.484.2047K}; \citealt{2019MNRAS.486.3892R}) and observations (\citealt{2018ApJ...861...50B}; \citealt{2018MNRAS.476..979P}; \citealt{2019ApJ...884..180D}; \citealt{2019MNRAS.488..685M}) are starting to find that AGN feedback can be equally, or even more, significant than SN feedback in dwarf galaxies.

\section{AGN vs SN feedback in dwarf galaxies}
In cosmological simulations AGN feedback is a crucial ingredient in order to reproduce the observed properties of massive galaxies and the galaxy luminosity function, found to break at $L_{*}$ (e.g. \citealt{2006MNRAS.365...11C}; \citealt{2017MNRAS.465...32B}; \citealt{2017ApJ...844...31C}). AGN feedback is also required to explain the baryon cooling efficiency of massive galaxies, while in the low-mass regime SN feedback is sufficient (\citealt{2013ApJ...770...57B}). The change of slope from SN- to AGN-regulated regimes is found to occur at $L_{*}$, or at $M_{*} \sim 3 \times 10^{10}$ M$_{\odot}$, close to the transitional mass typically used to distinguish between massive and dwarf galaxies. 
Observational evidence for AGN feedback regulation of massive galaxies comes from the spatial coincidence between the large-scale X-ray cavities of galaxy clusters and the radio jets of their central supermassive black holes (e.g. \citealt{2000MNRAS.318L..65F}; \citealt{2000ApJ...534L.135M}; \citealt{2012MNRAS.421.1360H}) and from the finding that the star-formation rate of local massive galaxies depends on supermassive black hole mass (\citealt{2018Natur.553..307M}). The no dependence of star-formation rate with black hole mass for local dwarf galaxies hosting low-mass AGN was instead taken as evidence for SNe being the dominant source of feedback governing such galaxies (\citealt{2018ApJ...855L..20M}) as so far assumed in most numerical simulations. This was reinforced by the finding that the $M_\mathrm{BH}-\sigma$ correlation changed its slope when moving to the low-mass end (i.e. below stellar velocity dispersion $\sigma \sim$ 100 km s$^{-1}$) at a transitional stellar mass of $M_{*} \sim 5 \times 10^{10}$ M$_{\odot}$ that was fully consistent with that of the break in the galaxy luminosity function and change of regimes in baryon cooling efficiency (\citealt{2018ApJ...855L..20M}).

However, independent studies performed at the same time indicated opposite results: simulations show that the stellar debris from tidal disruption events could fuel and grow seed black holes (\citealt{2017NatAs...1E.147A}; \citealt{2019MNRAS.483.1957Z}; \citealt{2020arXiv200308133P}) whose feedback could become relevant and have significant effects on the host galaxy (\citealt{2019MNRAS.483.1957Z}). Observationally, long-slit and integral-field unit spectroscopy studies of two different samples of quiescent dwarf galaxies revealed that they possibly host AGN, which could be preventing the formation of stars in such galaxies (\citealt{2019ApJ...884..180D}; \citealt{2018MNRAS.476..979P}). AGN feedback could also explain the finding, based on HI observations, of a sample of gas-depleted isolated dwarf galaxies possibly hosting AGN (\citealt{2018ApJ...861...50B}). X-ray or radio observations are however required to confirm the presence of AGN in these quiescent dwarf galaxies. Based on deep radio observations of the COSMOS field (\citealt{2017A&A...602A...1S}), \cite{2019MNRAS.488..685M} found a sample of radio AGN dwarf galaxies whose radio jets have powers and efficiencies as high as those of massive galaxies. This indicates that AGN feedback could be as significant in dwarf galaxies as in more massive ones. In massive galaxies AGN feedback can both prevent and trigger star formation on pc or kpc scales around the black hole (e.g. \citealt{2013ApJ...772..112S}; \citealt{2016A&A...593A.118Q}; \citealt{2017Natur.544..202M}), which can affect the amount of material available for the black hole to grow. If AGN feedback is also significant in dwarf galaxies, it could be that seed black hole growth is not hampered by SN feedback but enhanced by AGN feedback (\citealt{2019NatAs...3....6M}).

\section{Dwarf galaxy mergers}
Dwarf galaxy mergers are another factor to be taken into account. Some AGN are found in dwarf galaxies undergoing a merger (e.g. \citealt{2013MNRAS.435.2335B}; \citealt{2017ApJ...836..183S}) and several IMBH candidates are located in the outskirts of large galaxies, which suggests they are the nucleus of a dwarf galaxy stripped in the course of a merger (e.g. \citealt{2009Natur.460...73F}; \citealt{2013MNRAS.436.1546M,2013MNRAS.436.3128M,2015MNRAS.448.1893M,2018MNRAS.480L..74M}). Such minor mergers are expected to be very common and to trigger up to 50 \% of the local star formation activity (\citealt{2014MNRAS.437L..41K}). Cosmological simulations show that major mergers of dwarf galaxies can nonetheless also be very frequent (\citealt{2010MNRAS.406.2267F}; \citealt{2014ApJ...794..115D}). Studies of individual systems (e.g. \citealt{2015AJ....149..114P,2020AJ....159..141P}) and large surveys (e.g. \citealt{2015ApJ...805....2S}; \citealt{2018ApJS..237...36P}) indeed show that dwarf galaxies can be commonly found as interacting or merging pairs. If in the course of such dwarf-dwarf galaxy mergers the two IMBHs coalesce and high accretion rates are triggered, the resulting black hole could have a mass significantly enhanced with respect to that of the initial seeds (\citealt{2014ApJ...794..115D}; \citealt{2019NatAs...3....6M}). 

The finding of low-mass AGN in such dwarf-dwarf galaxy mergers is however scarce. \cite{2014ApJ...787L..30R} find an AGN with $M_\mathrm{BH} \sim10^5-10^7$ M$_{\odot}$ in the southern member of the dwarf galaxy pair Mrk709, but no black hole is detected for the northern galaxy. Statistically, 9\% of the low-mass AGN in \cite{2011ApJ...742...68J} are found in dwarf galaxies with a possible companion and eight out of the 23 (i.e. 35\%) AGN dwarf galaxies at $z<$0.3 of \cite{2018MNRAS.478.2576M} seem also to have a companion or to be undergoing a merger (see Fig. 1). However, in all these systems the stellar mass of the companion galaxy and whether it hosts a low-mass AGN is unknown. Whether dual AGN are formed or AGN activity is triggered during the merger of two dwarf galaxies is thus far from clear. Even if a dual low-mass AGN was formed, it is yet unclear whether the merger of the two IMBHs would occur as in dwarf galaxies dynamical friction might not be efficient enough to remove the necessary angular momentum to form a close black hole binary so that the black holes might stall and not merge (\citealt{2018ApJ...864L..19T}). Further studies are thus required to probe the role of dwarf galaxy mergers in seed black hole growth.

\section{Conclusions}
A myriad of low-mass AGN are being found in dwarf galaxies both in the local Universe and at the peak of cosmic star formation history. Such low-mass AGN could host the ungrown leftover of the first seed black holes formed in the early Universe and invoked to explain the rapid growth of supermassive black holes by $z\sim$7. AGN feedback, tidal disruption events, and dwarf galaxy mergers can nonetheless yield significant growth of these primordial seeds, in which case local low-mass AGN in dwarf galaxies should not be considered the untouched relics of the high-$z$ seed black holes. This has crucial implications not only for seed black hole formation models, but also for understanding the mechanisms governing black hole-galaxy evolution in the realm of dwarf galaxies. 

\begin{figure}
\includegraphics[width=\textwidth]{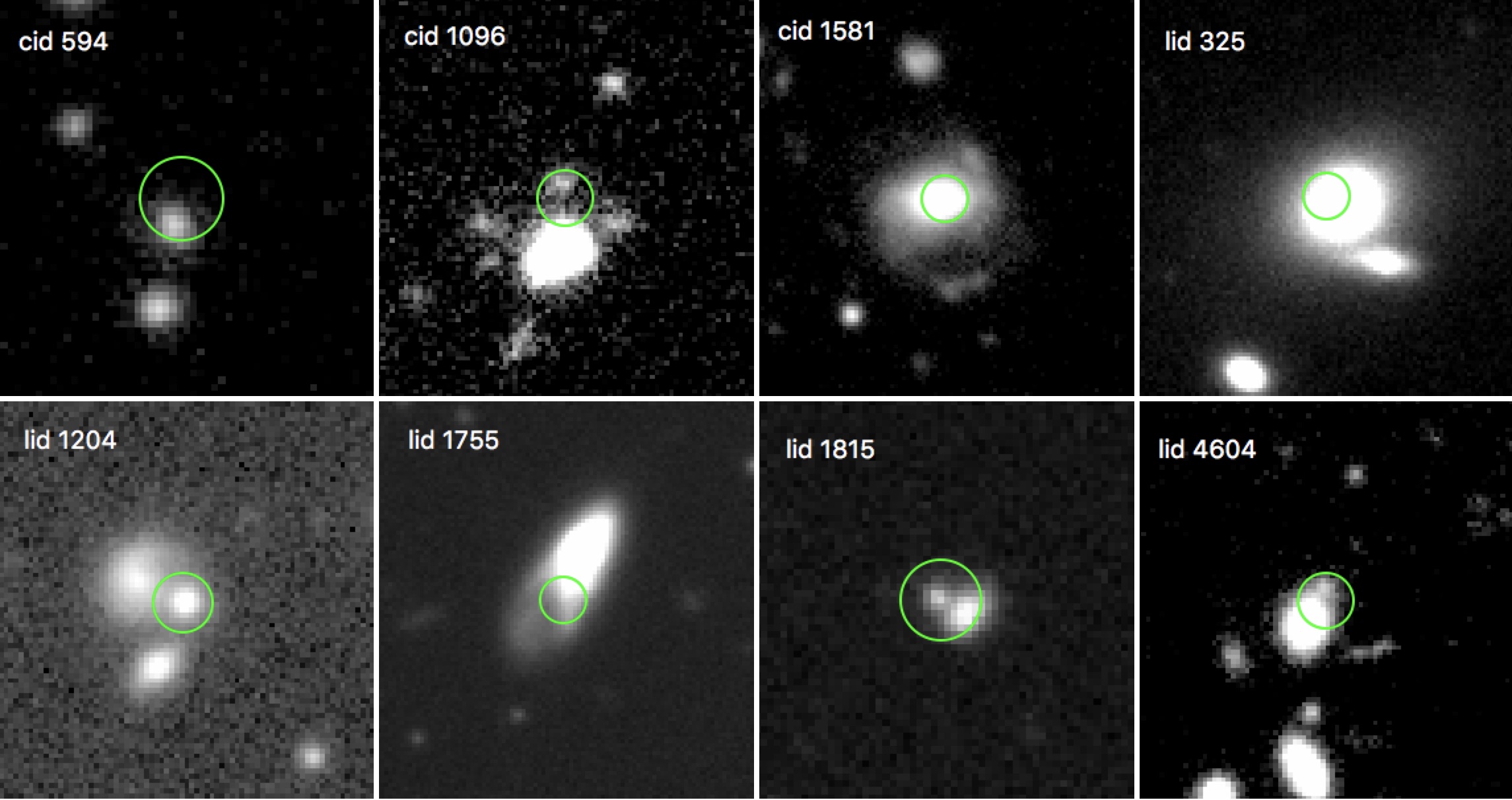}
\caption{Subaru Hypercam images of eight of the 23 X-ray AGN dwarf galaxies at $z<$0.3 of \cite{2018MNRAS.478.2576M} showing possible companions or possibly undergoing a merger. The \textit{Chandra} X-ray position is marked with a green circle of radius 1 arcsec.}
\end{figure}

\bibliographystyle{mnras}
\bibliography{/Users/mmezcua/Documents/referencesALL}

\end{document}